\documentclass[aps,prd,showpacs,amsmath,amssymb]{revtex4}

\begin{document}


\title{Comment on ``Bayesian Analysis of Pentaquark Signals from CLAS
Data'', with Response to the Reply by Ireland and Protopopsecu}

\author{Robert D. Cousins}
\affiliation{%
Department of Physics and Astronomy, University of California,
Los Angeles, CA 90095}%
\email{cousins@physics.ucla.edu}

\date{August 22, 2009}

\begin{abstract}
The CLAS Collaboration has published an analysis using Bayesian model
selection.  My Comment criticizing their use of arbitrary prior
probability density functions, and a Reply by D.G. Ireland and
D. Protopopsecu, have now been published as well.  This paper responds
to the Reply and discusses the issues in more detail, with particular
emphasis on the problems of priors in Bayesian model selection.
\end{abstract}

\maketitle

\section{Introduction}
The CLAS Collaboration \cite{clas08}, citing the pioneering work of
Harold Jeffreys \cite{jeff61}, has published a Letter \cite{clas08}
that claims to illustrate a general method that ``could be applied to
any data set where a search for a new state has been carried out'',
providing a ``quantitative measure'' for judging potential discovery
results by using the formalism of Bayesian model selection.  My
Comment \cite{comment} criticizing their use of arbitrary prior
probability density functions, and a Reply \cite{reply} by
D.G. Ireland and D. Protopopsecu, have now been published as well.  As
I believe that the Reply did not satisfactorily address the points
raised by my Comment, and that it was not informed by other points in
the articles I cited, I elaborate in this post to the arxiv.

All Bayesian calculations, and in particular model selection results,
are potentially sensitive to the choice of prior pdf.  My short
Comment \cite{comment}, reproduced in Sec.~\ref{comment}, focused on
the statement in the Letter \cite{clas08}: ``We assume that each prior
is a uniform distribution between a lower and upper limit since this
represents the least initial bias.'' This statement goes against the
entire thrust of Jeffreys' book and subsequent research: Jeffreys
explains in convincing detail the contradictions one reaches by such
use of Laplace's idea of a uniform prior.  Of course, a lot has been
learned in the nearly half century since Jeffreys' last edition
appeared in 1961, and my Comment included references to some more
recent key papers and discussion, notably by Jos\'e Bernardo and Jim
Berger and collaborators, and the review by Kass and Wasserman.

My Comment focused on problems in prior specification which are
present already in Bayesian parameter estimation. In Bayesian Model
Selection, there is an additional major concern arising from computing
a ratio of two integrals that are evaluated in parameter spaces of
different dimensionality.  In the CLAS Collaboration's Letter
\cite{clas08}, one has a four-dimensional prior in the denominator,
while the prior in the numerator has the same four dimensions plus
three additional ones.  Specification of the prior in these extra
dimensions without arbitrarily affecting the answer is a difficult
problem which is not addressed by the uniform priors described in
the Letter.

To spell out these objections further, in Sec.~\ref{reply} I respond
to some specific statements in the Reply \cite{reply}.
Sec.~\ref{truncation} explains why the evidence ratio in
the Letter has a steep dependence on the arbitrarily-chosen
upper and lower limits referred to in the above quote. Indeed the
method advocated in the Letter, in which ``The prior parameter ranges
were established by performing an initial fit and setting the limits
to be $\pm$50\% of the values found'' \cite{clas08}, is of a type
cautioned against in the Bayesian literature.  I conclude in
Sec.~\ref{conclusion}.

\section{My Comment as Published in PRL \cite{comment}}
\label{comment}

The CLAS collaboration, in presenting a Bayesian analysis of two
searches for pentaquarks \cite{clas08}, suggests ``an alternative
means of quantifying the evidence for discovery. What is specifically
required is a quantitative comparison between the two
hypotheses\dots", and concludes, ``More generally, this method could
be applied to any data set where a search for a new state has been
carried out, and can provide a quantitative measure with which to
judge whether or not a result represents a discovery."  Especially
given the repeated emphasis on ``quantitative", one must challenge the
statement regarding the prior probability densities that CLAS used for
the continuous parameters $\xi$: ``We assume that each prior is a
uniform distribution between a lower and upper limit since this
represents the least initial bias."  This assumption of uniform priors
for up to seven parameters in arbitrary metrics (while also not
defining ``bias") is in conflict with the large literature on priors
(both subjective and non-subjective), including the influential work
of Jeffreys \cite{jeff61} cited by Ref.~\cite{clas08} as the basis of
their method.

For a continuous parameter $\xi$ with reasonably behaved Bayesian
prior probability density function $P(\xi|M)$, one can use the
probability integral transformation \cite{pearson38} to construct a
function $\zeta(\xi)$ for which the $P(\zeta|M)$ is uniform.  Thus the
choice of $P(\xi|M)$ is equivalent to the choice of the metric
$\zeta(\xi)$ in which $P$ is uniform; arbitrarily assuming $\zeta(\xi)
= \xi$ as implicit in Ref.~\cite{clas08} is without justification.  In
parameter estimation problems, Jeffreys advocated a general rule for
non-subjective priors using the Fisher information, although he saw
the need for modification in cases of multiple parameters.  Advocates
of so-called ``objective Bayesianism", Bernardo \cite{bernardo79} with
Berger \cite{bergerbern89,bergerbern92}, and others then developed
``reference priors" for multiple parameters.  For model selection,
other considerations may lead to yet other priors
\cite{jeff61,bergerpericchi}.

The many issues of priors selected by such ``formal rules" are
reviewed by Kass and Wasserman \cite{kass96} and by Berger and
Pericchi \cite{bergerpericchi}. Bernardo discusses his views in
Ref.~\cite{bernardo97}, with commentary from statisticians.  Other
spirited discussion follows articles by Berger \cite{berger06} and by
Goldstein \cite{goldstein06} who advocate, respectively, the {\em
objective} and {\em subjective} Bayesian approaches.  Efron has also
noted, \cite{efron03}, ``Perhaps the most important general lesson is
that the facile use of what appear to be uninformative priors is a
dangerous practice in high dimensions."

For Bayesian analyses relevant to a community of scientists one should
study the {\it sensitivity} of the result to variation of the priors.
Goldstein \cite{goldstein06} says, ``We can then produce the range of
posterior judgements, given the data, which correspond to the range of
`reasonable' prior judgements held within the scientific community."
Bernardo \cite{bernardo97} says, ``Non-subjective Bayesian analysis is
just a part -- an important part, I believe -- of a healthy {\em
sensitivity analysis} to the prior choice\dots'' Berger also argues
that objective priors such as Jeffreys' priors and their
generalizations lead to good frequentist properties which are welcomed
by many statisticians although not part of the Bayesian paradigm.

In summary, Ref.~\cite{clas08} does not justify its assumption of
uniform priors in the chosen parameter metrics $\xi$.  A ``healthy''
sensitivity analysis should accompany any use of this method for
scientific communication, but is absent in Ref.~\cite{clas08}.
Following prominent objective Bayesians \cite{berger06,bernardo97}, it
would also be useful to understand the frequentist properties of the
method in order to facilitate comparison of results from different
paradigms.

\section{Response to the Reply to My Comment}
\label{reply}

I do not believe that the Reply \cite{reply} by Ireland and
Protopopsecu adequately addresses my criticisms with respect to the
priors.  There is also no study presented of the frequentist sampling
properties of their result.  Thus the scientific conclusions one
obtains by following the general method as presented in the Letter
\cite{clas08} are without proper foundation, and cannot be interpreted
in the ``quantitative'' fashion claimed in the Letter.

Below are some italicized quotes from the published Reply and my
further comments.  \\

{\it ``The choice of a Gaussian function to represent a peak is a
standard one, since the three parameters required to specify it are
related to quantities with physically meaningful interpretations:
centroid position, detector resolution and signal strength. The limits
on the possible value of centroid position correspond to the mass
range for which the experiment is sensitive, so it is a {\rm location}
parameter for which a uniform prior is a reasonable choice.''}  \\

In this (August 2009) version of this note, I modify my comment on this
point.  My earlier comment contained the following two paragraphs:

``The fact that a parameterization is a standard representation in
physics formulas has nothing to do with whether or not the those are
the metrics in which the prior should be flat.  The claim that mass is
a location parameter for which a uniform prior is reasonable is false.
The prototypical example that Jeffreys used to show the non-universal
applicability of uniform priors is in fact a physical parameter with a
semi-infinite range of possible values: the charge of the electron
(pp. 104-105 in the 2nd edition, pp. 119-120 in the third edition).
His analysis applies equally to mass.  Those pages are in the
part of the book on estimation, and as noted in my Comment, one can be
led to different priors in model selection; but in neither case is
mass a location parameter with uniform prior.''

``The classic example of a location parameter is the mean $\theta$ of a
Gaussian where $\theta$ is in $(-\infty,\infty)$.  But taking a {\it
semi}-infinite physical parameter such as mass and smearing it with a
Gaussian resolution function does not turn it into a location
parameter.''

I now think the issue is not as clear-cut as I portrayed it, for very
interesting reasons.  The above argument by Jeffreys is an argument
based on considerations of the physical parameter itself.  This is
{\it in contrast} to the argument (in the same book) leading to
Jeffreys's Rule for priors that are known as ``Jeffreys priors'', and
that are generalized to the Reference Priors of Bernardo and
collaborators.  Jeffrey's Rule for priors is completely based on the
{\it measurement} process, i.e., the specific experimental setup one
uses in order to ``make the measurement''; this is what statisticians
typically call ``the model''. This is distinct from considerations
(such as the scaling argument for the electron charge) about the
physical parameter itself!  That is why, for instance, the Jeffreys
prior for the binomial parameter is different in the binomial model
than in negative binomial model, in violation of the strong likelihood
principle, even if it is the same physical parameter. So when using
Jeffrey's Rule, it is indeed the case that simply measuring a
parameter with a Gaussian resolution function implies a uniform prior,
independent of one's notions about the physical parameter itself!

As to the issue of a restriction on allowed values of the parameter,
Berger \cite{decision} addresses exactly this point:

``One important feature of the Jeffreys noninformative prior is that
it is not affected by a restriction on the parameter space.  Thus, if
it is known in Example 5 [in which $\theta$ is a location parameter]
that $\theta>0$, the Jeffreys noninformative prior is still
$\pi(\theta)=1$ (on $(0,\infty)$, of course).  This is important,
because one of the situations in which noninformative priors prove to
be extremely useful is when dealing with restricted parameter spaces
(see Chapter 4).  In such situations we will, therefore, simply assume
that the uninformative prior is that which is inherited from the
unrestricted parameter space.''

My above-quoted comments on this point were thus wrong from the point
of view of objective Bayesianism; the scaling argument of Jeffreys
which I quoted is eclipsed by Jeffreys's Rule.  On the one hand this
might seem to be a compelling argument against such ``objective''
priors from Jeffreys's Rule, but on the other hand it is worth
recalling arguments in favor of such objective priors, namely that they
``let the data speak the loudest'', and that since the time of Welch
and Peers they are associated with probability matching to a certain
order.

This discussion seems to make quite clear how so-called objective
priors are different from subjective priors.  Subjective priors encode
an individual's prior beliefs about the parameter, while objective
priors encode properties of the measuring apparatus (including the
stopping rule, in violation of the likelihood principle).  The
arguments I quoted advocating a sensitivity analysis would seem to
take on even greater force in these circumstances!  \\

{\it ``The width and height of the Gaussian are, strictly speaking,
{\rm scale} parameters for which a Jeffreys' prior may be
appropriate. However, for the calculation of evidence integrals, we
require normalized priors, so a Jeffreys' prior [$f(x)\propto 1/x$]
needs to be normalized between two limits. Limits on the width of the
peak are naturally suggested: the minimum by detector resolution and a
maximum such that a peak is not confused with a background shape.''}
\\

The uniform prior is also unnormalizable over all space.  The authors
do not explain why they chose a truncated uniform prior rather than a
truncated version of some other unnormalizable prior, if that was the
concern.  But the larger issue is that if one is in a situation where
truncation is needed in order to make the prior normalizable, then the
result of the model selection calculation will depend on the endpoints
used for the truncation.  I come back to this crucial point in
Sec.~\ref{truncation} below.  \\

\noindent {\it ``For the signal parameter, we must include zero, since
the posterior probability density function shows that, even in cases
where the maximum likelihood is non-zero (i.e. a peak is most likely),
there is still a significant probability that the results are
consistent with zero signal. The Jeffreys' prior is undefined at
$x=0$, so an alternative could be to use a gamma distribution, which
is normalized, has a similar decay as $x\rightarrow\infty$, and is
defined as $x\rightarrow0$.}  \\

The Bayesian literature on model selection (starting with Jeffreys)
has a lot of discussion about the situation, as we have here, where
the crucial scientific question is whether or not a particular
parameter is zero (i.e., corresponding to the classical case of a
point null hypothesis).  There are various ways to approach this, for
example concentrating some prior in a delta function at the point, and
spreading out the rest.  (This is already in Jeffreys' 2nd edition.)
A more comprehensive look at the professional literature is needed to
understand the subtleties which have been ignored here.  \\

\noindent {\it ``We studied the problem with these alternative priors,
and found no significant difference in the results obtained; the use
of uniform priors was thus motivated by using the simplest form
appropriate to the problem.''}  \\

Without a quantitative exposition or knowing the scope of the
sensitivity analysis, one cannot evaluate the claim that their method
provides a ``quantitative'' result.  Were the limits of the mass range
changed?  What range of shapes of gamma functions was explored?  \\

{\it ``In addition, the sensitivity of the results to the parameters
of the background is minimal. The evidence ratios (or Bayes Factors)
are in the form of (logarithms of) ratios of integrals. Any slowly
varying prior in the space of the background parameters will thus
result in approximately constant factors that cancel.''}  \\

Again, whether something is ``minimal'' or ``approximately constant''
is in the eye of the beholder, so the authors should provide the list
of alternatives explored and the numerical results.  \\

{\it ``In the calculations, the likelihood functions for each data
model must be integrated over all parameter space. The effect of
priors is to weight the likelihood functions. In practice, the
likelihood functions for the study in the letter are significantly
non-zero in only a small region of parameter space. It would thus take
a rapidly varying prior in this region to make a noticeable difference
to the integrals, and one would of course have to justify this rapidly
varying prior.''}  \\

A rapidly varying prior in their chosen metric will be a uniform prior
in another metric; what is the criterion for choosing the preferred
metric?  As I discuss below, the location of the endpoints of the
uniform prior can affect the answer, even if the likelihood is zero
near the endpoints.

With seven parameters, the authors may be surprised to find that the
volume effect (most of the prior probability located near the boundary
of their 7-D hypercube) distorts the posteriors.  \\

{\it We used the term "least bias" to indicate that, as far as
possible, we wanted to see what information one could extract from the
data, whilst introducing only minimal prior prejudice.}\\

As shown by Jeffreys \cite{jeff61} and reviewed by Kass and Wasserman
\cite{kass96}, using uniform priors in the manner of the Letter
\cite{clas08} does {\it not} satisfy this desire.  This is also
extensively discussed by Bernardo in the ``dialogue'' cited in my
Comment \cite{bernardo97}.\\

\noindent {\it How one achieves this is an open question, and it
should be noted that the debate within the statistics community
appears far from settled, as attested by the papers cited in the
Comment [2,3], and subsequent contributions in the same publication.}
\\

This was the reason I wrote the Comment: The Letter's statement that
their uniform priors had the ``least bias'' was without foundation.
It appears that the authors of the Reply now agree.\\

{\it ``To conclude, the use of alternative priors makes little
difference to the results of our study. The measured data in this case
therefore contain sufficient information to dominate the calculation
of the probabilistic quantities of interest.''}  \\

The reader should be provided with details of the alternative priors
and numerical results in order to judge what ``little difference''
means, since Sec.~\ref{truncation} below indicates that this is not
the case.  \\

\noindent {\it Actual claims of discovery would require a more
detailed examination of evidence than presented in the Letter. We
therefore fully agree with the author of the Comment that, in general,
a sensitivity analysis of results to the choice of priors (and data
models) is essential.''}  \\

The main scientific point of the Letter \cite{clas08} was to claim
that the first CLAS result was actually ``inconclusive'' (and if
anything weak evidence against a peak), contradicting CLAS's earlier
published claim of 5.2$\sigma$ ``observation'' of a peak.  It would
seem that such an extraordinary situation should be backed up with the
sort of ``detailed examination'' that the Reply authors agree is
required for a discovery. This should include the dependence on the
limits (endpoints) of the prior as discussed in Sec.~\ref{truncation}
below.

\section{The Problem of Differing Dimensions in Model Selection and the
Dangers of Arbitrarily Truncating the Prior}
\label{truncation}
A difficulty in the present model selection calculation is a common
one in Bayesian analysis: Model A has four parameters, and Model B has
the same four parameters, plus three more.  If the priors for the
three additional parameters are unnormalizable (e.g. a uniform prior
extending to $\infty$), then the answer is completely arbitrary:
multiplying the prior by a constant in the numerator but not the
denominator (where it does not appear) will change the evidence ratio.
The Reply alludes to this, and in the Letter the uniform prior was
truncated at a location well outside the range where likelihood
function is non-negligible.  This truncation is perhaps
plausible-sounding since such a truncation may be made without severe
effect in estimation problems. However, the effect in model selection
is unfortunately that which often occurs in physics when dealing with
an infinity by introducing a cutoff: one simply replaces the infinity
with a dependence on the cutoff, which is arbitrary.

The Letter set the endpoints to be $\pm$50\% of the best-fit values.
(``The prior parameter ranges were established by performing an
initial fit and setting the limits to be $\pm$50\% of the values
found'' \cite{clas08}.) The Reply \cite{reply} appears to contradict
this statement from the Letter, by saying that zero was included in
one interval; but as this statement in the Letter has not been
explicitly retracted, we can take it as the generally applicable
method advocated by the CLAS Collaboration.  Then let $f$ denote this
percentage, i.e. $f=0.5$ in the Letter, and let $\hat\xi$ denote the
best-fit value of a parameter $\xi$.  Then the prior is:
\begin{equation}
P(\xi|M) = \left\{ \begin{array}{ll}
   1/(2f\hat \xi), 
            & \mbox{$\hat\xi-f\hat\xi < \xi < \hat\xi + f\hat\xi$}\\
   0, & \mbox{otherwise.}
                   \end{array}
           \right.
\end{equation}
Thus, even if the likelihood functions are all negligible near the
endpoints of the uniform prior, after all integrations are performed
the arbitrary factor $(1/f\hat\xi)$ will appear in the numerator of
the evidence ratio, and not be canceled be a factor in the
denominator.  {\it Thus the arbitrary constant of an unnormalizable
uniform prior is just replaced by an arbitrary constant determining
the height of the normalized uniform prior (via the arbitrary
specification of the width and the normalization condition).}

As the Letter has one such factor $f$ for each of its three extra
parameters in the numerator, the evidence ratio goes as $f^3$.  That
is, if $f$ is varied from 25\% to 75\%, the evidence ratio changes by
a factor of 27, and its logarithm, $\ln(R_E)$, changes by 3.3 units.
This is enough to change the evidence by two categories of strength in
Jeffrey's scale!

The value of $f$ used for the mass parameter deserves special
attention, since it controls the ``Occam's razor'' effect due to where
one is looking in mass: a firm prediction of the mass of the
pentaquark, followed by a peak at that location, should result in an
enhanced evidence ratio resulting from a small value $f$.  It is not
clear to what extent the analysis in the Letter in fact {\em reduced}
the evidence ratio derived from the first data set by using an
inflated value of $f$.

Given such strong dependence on an arbitrary parameter $f$, it is hard
to comprehend the claim in the Reply that a sensitivity analysis was
performed with ``no significant difference in the results obtained''.

(The Letter was not clear as to whether or not the limits on the four
parameters in the 3rd-order polynomial were the same in the numerator
as in the denominator.  If different ``values found'' by the fit led
to different limits in the numerator and denominator, then this adds
to the non-canceling dependence on $f$.)

This issue is of course known in the Bayesian literature that I cited
in my Comment.  For example, Berger and Pericchi \cite{bergerpericchi}
list ``Difficulty 3. Use of vague proper priors usually give bad
answers in Bayesian model selection'', with a specific example where
the Bayes factor depends the arbitrarily chosen ``large'' value of
$K$, which in their example sets the size of the region over which the
prior is appreciable.  They conclude, ``The short story is never use
vague priors for model selection...'' (and in fact prefer an improper
prior in that particular example). The limited-length uniform priors
used in the Letter suffer from the same disease of the Bayes factor
depending on the arbitrary $f$.

In his article in Bayesian Analysis cited by my Comment, Jim Berger
\cite{berger06} has a section entitled ``Dangers of Casual Objective
Bayesian Analysis'' which is even more explicit in urging caution in
use of ``{\em pseudo-Bayes}'' analyses: ``...while they utilize
Bayesian machinery, they do not carry with them any of the guarantees
of good performance that come with either true subjective analysis
(with a very extensive elicitation effort) or (well-studied) objective
Bayesian analysis...and hence must be validated by some other route.''
He specifically warns in a section entitled ``Truncation of the
parameter space'' that truncation at large $\pm K$ to avoid having an
improper pdf must be done with care: ``At the very least, this
approach should only be used if a very careful sensitivity study is
done with respect to these bounds (and with bounds for different
parameters varying independently in the sensitivity study.''  The
context of these quotes is in terms of avoiding an improper posterior
by truncating an improper prior, but the concern is exactly paralleled
in truncation of the prior in the model selection problem of
the Letter \cite{clas08}.  

The last subsection in Berger's section on 
``Dangers of Casual Objective Analysis'' is worth quoting extensively:
\cite{berger06}:

``{\bf Data-dependent vague proper priors.} The second common
data-dependent procedure is to choose priors that span the range of
the likelihood function. For instance, one might choose a uniform
prior over a range that includes most of the mass of the likelihood
function, but that does not extend too far (thus hopefully avoiding
the problem of using a `too vague' proper prior). Another version of
this procedure is to use conjugate priors, with parameters chosen so
that the prior is spread out somewhat more than the likelihood
function, but is roughly centered in the same region. The two obvious
concerns with these strategies are that (i) the answer can still be
quite sensitive to the spread of the rather arbitrarily chosen prior;
and (ii) centering the prior on the likelihood is a quite problematic
double use of the data. Also, in problems with complicated
likelihoods, it can be very difficult to implement this strategy
successfully.\dots In conclusion, while these pseudo-Bayesian
techniques can be useful as data exploration tools, they should not be
confused with formal objective Bayesian analysis, which has very
considerable extrinsic justification as a method of analysis.''

\section{Conclusion}
\label{conclusion}
If we are going to use Bayesian techniques in our research, then we
should read and understand a representative sampling of the relevant
Bayesian literature.  I urge anyone contemplating an objective
Bayesian analysis to read Kass and Wasserman \cite{kass96} before
attempting to write down a so-called objective or noninformative prior
in a desire to ``represent the least initial bias''.  If one wants to
go beyond parameter/interval estimation and get into model selection,
the article by Berger and Pericchi \cite{bergerpericchi}, also cited
in my Comment, is a must for beginning to appreciate the difficulty of
the subject and potential pitfalls; the volume has other valuable
articles and commentary as well.  For another perspective and pointers
to a much broader discussion, see Chapter 6 (including the
``Bibliographic Note'' in Sec. 6.9) of the text by Gelman et
al. \cite{gelman}.  Kass and Rafferty give another brief synopsis in
Sec. 5.1 of their article on Bayes Factors ~\cite{kassraftery}.  Of
course Jeffreys' classic monograph also still provides insightful
reading and historical perspective.

The articles by Berger and by Goldstein and the ensuing discussion in
Bayesian Analysis \cite{berger06,goldstein06} are a great introduction
to the discussion within the Bayesian community.  While there is quite
a spirited discussion, it is clear that there is a consensus
recommendation for a ``healthy'' sensitivity analysis in any Bayesian
analysis used for scientific communication.  In a Model Selection
analysis, particular caution is needed when using priors in which
arbitrary constants in the normalization do not cancel in the evidence
ratio.  The method advocated by the CLAS Collaboration \cite{clas08},
while applying the established Bayesian model selection formalism,
used such arbitrary inputs and thus the ``quantitative'' output should
also be regarded as arbitrary, until a ``healthy'' sensitivity
analysis is displayed, and/or the sampling properties are understood.

\begin{acknowledgments}
This work was supported by the U.S. Department of Energy.
\end{acknowledgments}


\begin{thebibliography}{12}
\expandafter\ifx\csname natexlab\endcsname\relax\def\natexlab#1{#1}\fi
\expandafter\ifx\csname bibnamefont\endcsname\relax
  \def\bibnamefont#1{#1}\fi
\expandafter\ifx\csname bibfnamefont\endcsname\relax
  \def\bibfnamefont#1{#1}\fi
\expandafter\ifx\csname citenamefont\endcsname\relax
  \def\citenamefont#1{#1}\fi
\expandafter\ifx\csname url\endcsname\relax
  \def\url#1{\texttt{#1}}\fi
\expandafter\ifx\csname urlprefix\endcsname\relax\def\urlprefix{URL }\fi
\providecommand{\bibinfo}[2]{#2}
\providecommand{\eprint}[2][]{\url{#2}}

\bibitem[{\citenamefont{{D.G. Ireland {\em et al.}}}(2008)}]{clas08}
\bibinfo{author}{\bibnamefont{{D.G. Ireland {\em et al.}}}}
  (\bibinfo{collaboration}{CLAS Collaboration}), 
``A Bayesian analysis of pentaquark signals from CLAS data'',
\bibinfo{journal}{Phys. Rev. Lett.}
  \textbf{\bibinfo{volume}{100}}, \bibinfo{pages}{052001}
  (\bibinfo{year}{2008}), \eprint{arXiv:0709.3154 [hep-ph]}.

\bibitem[{\citenamefont{Jeffreys}(1961)}]{jeff61}
\bibinfo{author}{\bibfnamefont{H.}~\bibnamefont{Jeffreys}},
  \emph{\bibinfo{title}{Theory of Probability}} (\bibinfo{publisher}{Oxford
  University Press}, \bibinfo{address}{New York}, \bibinfo{year}{1961}),
  \bibinfo{edition}{3rd} ed.

\bibitem{comment}
Robert D. Cousins, Phys. Rev. Lett. {\bf 101} 029101 (2008).

\bibitem{reply}
D. G. Ireland and D. Protopopescu,
Phys. Rev. Lett. {\bf 101} 029102 (2008).

\bibitem[{\citenamefont{Pearson}(1938)}]{pearson38}
\bibinfo{author}{\bibfnamefont{E.~S.} \bibnamefont{Pearson}},
``The Probability Integral Transformation for Testing Goodness 
of Fit and Combining Independent Tests of Significance'',
  \bibinfo{journal}{Biometrika} \textbf{\bibinfo{volume}{30}},
  \bibinfo{pages}{134} (\bibinfo{year}{1938}), \bibinfo{note}{and references
  therein to R. Fisher and K. Pearson.}
\url{http://www.jstor.org/stable/2332229}

\bibitem[{\citenamefont{Bernardo}(1979)}]{bernardo79}
\bibinfo{author}{\bibfnamefont{J.~M.} \bibnamefont{Bernardo}},
``Reference Posterior Distributions for Bayesian Inference'',
  \bibinfo{journal}{Journal of the Royal Statistical Society. Series B
  (Methodological)} \textbf{\bibinfo{volume}{41}}, \bibinfo{pages}{113}
  (\bibinfo{year}{1979}).
\url{http://www.jstor.org/stable/2985028}

\bibitem[{\citenamefont{Berger and Bernardo}(1989)}]{bergerbern89}
\bibinfo{author}{\bibfnamefont{J.~O.} \bibnamefont{Berger}} \bibnamefont{and}
  \bibinfo{author}{\bibfnamefont{J.~M.} \bibnamefont{Bernardo}},
``Estimating a Product of Means: Bayesian Analysis with Reference Priors'',
  \bibinfo{journal}{Journal of the American Statistical Association}
  \textbf{\bibinfo{volume}{84}}, \bibinfo{pages}{200} (\bibinfo{year}{1989}).
\url{http://www.jstor.org/stable/2289864}

\bibitem[{\citenamefont{Berger and Bernardo}(1992)}]{bergerbern92}
\bibinfo{author}{\bibfnamefont{J.~O.} \bibnamefont{Berger}} \bibnamefont{and}
  \bibinfo{author}{\bibfnamefont{J.~M.} \bibnamefont{Bernardo}},
``Ordered Group Reference Priors with Application to the Multinomial Problem'',
  \bibinfo{journal}{Biometrika} \textbf{\bibinfo{volume}{79}},
  \bibinfo{pages}{25} (\bibinfo{year}{1992}).
\url{http://www.jstor.org/stable/2337144}

\bibitem[{\citenamefont{Berger and Pericchi}(2001)}]{bergerpericchi}
\bibinfo{author}{\bibfnamefont{J.}~\bibnamefont{Berger}} \bibnamefont{and}
  \bibinfo{author}{\bibfnamefont{L.}~\bibnamefont{Pericchi}}, 
``Objective Bayesian Methods for Model Selection: Introduction and 
Comparison'', in
  \emph{\bibinfo{booktitle}{Model Selection}}, edited by
  \bibinfo{editor}{\bibfnamefont{P.}~\bibnamefont{Lahiri}}
  (\bibinfo{publisher}{Inst. of Mathematical Statistics},
  \bibinfo{address}{Beachwood, Ohio}, \bibinfo{year}{2001}),
  vol.~\bibinfo{volume}{38} of \emph{\bibinfo{series}{Lecture Notes-Monograph
  Series}}, pp. \bibinfo{pages}{135--207}, \bibinfo{note}{with discussion}.
Vol. 38 at \url{http://www.imstat.org/publications/lecnotes.htm} has a link
to most of the text in Google Book Search.


\bibitem[{\citenamefont{Kass and Wasserman}(1996)}]{kass96}
\bibinfo{author}{\bibfnamefont{R.~E.} \bibnamefont{Kass}} \bibnamefont{and}
  \bibinfo{author}{\bibfnamefont{L.}~\bibnamefont{Wasserman}},
``The Selection of Prior Distributions by Formal Rules'',
  \bibinfo{journal}{Journal of the American Statistical Association}
  \textbf{\bibinfo{volume}{91}}, \bibinfo{pages}{1343} (\bibinfo{year}{1996}).
\url{http://www.jstor.org/stable/2291752},
\url{http://lib.stat.cmu.edu/~kass/papers/rules.pdf}

\bibitem[{\citenamefont{Irony and Singpurwalla}(1997)}]{bernardo97}
\bibinfo{author}{\bibfnamefont{T.~Z.} \bibnamefont{Irony}} \bibnamefont{and}
  \bibinfo{author}{\bibfnamefont{N.~D.} \bibnamefont{Singpurwalla}},
``Non-informative priors do not exist: A dialogue with Jose M. Bernardo'',
  \bibinfo{journal}{Journal of Statistical Planning and Inference}
  \textbf{\bibinfo{volume}{65}}, \bibinfo{pages}{159} (\bibinfo{year}{1997}),
\url{http://www.uv.es/~bernardo/Dialogue.pdf}

\bibitem[{\citenamefont{Berger}(2006)}]{berger06}
\bibinfo{author}{\bibfnamefont{J.}~\bibnamefont{Berger}},
``The Case for Objective Bayesian Analysis'',
  \bibinfo{journal}{Bayesian Analysis} \textbf{\bibinfo{volume}{1}},
  \bibinfo{pages}{385} (\bibinfo{year}{2006}), with discussion and
rejoinder.
\url{http://ba.stat.cmu.edu/vol01is03.php}

\bibitem[{\citenamefont{Goldstein}(2006)}]{goldstein06}
\bibinfo{author}{\bibfnamefont{M.}~\bibnamefont{Goldstein}},
``Subjective Bayesian Analysis: Principles and Practice'',
  \bibinfo{journal}{Bayesian Analysis} \textbf{\bibinfo{volume}{1}},
  \bibinfo{pages}{403} (\bibinfo{year}{2006}), with discussion and
rejoinder.
\url{http://ba.stat.cmu.edu/vol01is03.php}

\bibitem[{\citenamefont{Efron}(2003)}]{efron03}
\bibinfo{author}{\bibfnamefont{B.}~\bibnamefont{Efron}}, 
``Bayesians, Frequentists, and Physicists'', in
  \emph{\bibinfo{booktitle}{PHYSTAT2003: Statistical Problems in Particle
  Physics, Astrophysics, and Cosmology (SLAC, 8-11 Sep)}}, edited by
  \bibinfo{editor}{\bibnamefont{{L. Lyons, R. Mount, R. Reitmeyer}}}
  (Stanford Linear Accelerator Center, Menlo Park, 2003).
\url{http://www.slac.stanford.edu/econf/C030908/papers/MOAT003.pdf}

\bibitem{kassraftery}
Robert E. Kass and Adrian E. Raftery,
``Bayes Factors'', Journal of the American Statistical Association
{\bf 90}, 773 (1995).
\url{http://lib.stat.cmu.edu/~kass/papers/bayesfactors.pdf}.

\bibitem{gelman} 
A. Gelman, J.B. Carlin, H.S. Stern, and D.B. Rubin, {\em Bayesian Data
Analysis}, (Chapman \& Hall,Boca Raton, FL, 1998).

\bibitem{decision} 
James O. Berger, {\em Statistical Decision Theory and Bayesian
Analysis}, 2nd Ed., (Springer-Verlag, New York, 1985).  The quote is
from p. 89.


\end{thebibliography}



\end{document}